\documentclass[aps,prl,superscriptaddress,floatfix]{revtex4}

\usepackage{graphicx}
\usepackage{epsfig}
\usepackage[english]{babel}

\newcommand{\be}{\begin{equation}}
\newcommand{\bea}{\begin{eqnarray}}
\newcommand{\eea}{\end{eqnarray}}
\newcommand{\ee}{\end{equation}}

\def\one{\ensuremath{\hbox{$\mathrm I$\kern-.6em$\mathrm 1$}}}

\def\qed{\leavevmode\unskip\penalty9999 \hbox{}\nobreak\hfill
     \quad\hbox{\leavevmode  \hbox to.77778em{%
               \hfil\vrule   \vbox to.675em%
               {\hrule width.6em\vfil\hrule}\vrule\hfil}}
     \par\vskip3pt}
\newcommand{\beaa}{\begin{eqnarray*}}
\newcommand{\eeaa}{\end{eqnarray*}}
\newcommand{\bma}{\begin{subequations}}
\newcommand{\ema}{\end{subequations}}

\def\one{{\bf 1}}

\begin{document}

\title{Mapping local Hamiltonians of fermions to local Hamiltonians of spins}
\author{F. \surname{Verstraete}}
\affiliation{Institute for Quantum Information, Caltech, Pasadena, US.}
\author{J. I. \surname{Cirac}}
\affiliation{Max-Planck-Institut f\"ur Quantenoptik, Hans-Kopfermann-Str. 1,
  Garching, D-85748, Germany.}


\begin{abstract}
We show how to map local fermionic problems onto local spin problems on a lattice in any dimension. The main idea
is to introduce auxiliary degrees of freedom, represented by Majorana fermions, which allow us to extend the
Jordan-Wigner transformation to dimensions higher than one. We also discuss the implications of our results in
the numerical investigation of fermionic systems.
\end{abstract}


\maketitle

Many of the interesting strong correlation quantum phenomena that appear in Nature can be accounted for in terms
of discretized models, i.e., in lattices. The corresponding effective theories often deal either with fermions,
which can hop from one lattice site to another one as well as interact among them, or spins, which sit at
different lattice sites. For local Hamiltonians in one spatial dimension, the Jordan--Wigner transformation (JWT)
\cite{JW} allows us to map a {\em local} theory of fermions onto a {\em local} theory of spins, and thus those
systems exhibit related physical phenomena. In higher dimensions, however, this transformation gives rise to
non--local interactions between spins, and thus fermions and spins are expected to exhibit very different
behavior, at least when they are described in terms of local Hamiltonians.

In this paper we will show how one can always exactly map a local theory of fermions to the low energy sector of
a local theory of spins in a lattice of arbitrary dimension. To achieve this, we introduce extra degrees of
freedom in the form of Majorana fermions, which interact locally with the original ones, and which allow us to
perform a JWT in higher dimensions than one. Similar ideas were previously used by Bravyi and Kitaev
\cite{Bravyi} in the context of simulating fermionic quantum-computational  circuits with qubits; there they
showed how local fermionic gates can be  simulated with local gates on qubits if the quantum states are
rescricted  to live in a specific sector of the Hilbert space that is determined by  nonlocal gauge conditions.
Also, R. Ball \cite{Ball} has shown how these  gauge conditions can be transformed into local terms of the
Hamiltonian,  except at the boundary where nonlocal terms remain. We present a mapping which has the advantage
that the low energy sector is completely determined by local terms in the spin Hamiltonian. The spin Hamiltonians
obtained are exactly of the type studied by Wen et al. \cite{Wen1}, where the opposite question was studied, i.e.
how fermionic behavior could originate out of the low energy sector of a bosonic Hamiltonian.

The question analyzed in this paper is clearly of fundamental interest. However, it may also have important
implications in the numerical investigations of fermionic lattice problems, e.g. the Fermi--Hubbard model in two
dimensions. In fact, our original motivation was to develop a method to represent fermionic problems in terms of
local spin problems, so that we could use the algorithms we have recently developed to deal with those systems in
dimensions larger than one \cite{2D}. In that case, a direct numerical simulation in terms of the fermionic
degrees of freedom was not a good choice, since in Fock space one always obtains non--local terms (which are
directly related to the Jordan-Wigner transformation), and ground states of generic systems with long-range
interactions do not have to obey e.g. the area law \cite{Arealaw} and other characteristic properties that allow
for efficient simulations: ground states of local Hamiltonians are very special in the sense that they are
completely determined by their local reduced density operators, and hence a variational method will work fine if
it allows to approximate local properties well \cite{FC05}. This is one of the reasons why it is important that
the low energy sector of the spin Hamiltonian obtained is determined by local terms, and does not contain
nonlocal gauge conditions. Apart from that, our results may also prove useful for other standard simulation
methods, such as quantum Monte Carlo, since there one also effectively simulates spin Hamiltonians instead of
fermionic ones directly.

From the mathematical point of view, the natural formalism to describe a state of fermions is second
quantization. Within this language, the most general state of $N$ fermions can be parameterized as
 \[
 |\psi\rangle=\sum_{i_1,i_2,\cdots,i_N=0}^1C_{i_1,i_2,\cdots,i_N}
 \left(\hat{a}_1^\dagger\right)^{i_1}\left(\hat{a}_2^\dagger\right)^{i_2}
 \cdots\left(\hat{a}_N^\dagger\right)^{i_N}|\Omega\rangle
 \]
where $\hat{a}_i^\dagger$ are creation operators obeying the usual anticommutation relations, and where
$|\Omega\rangle$ denotes the vacuum. Note that, from the moment we consider second quantization, we have to
enumerate the fermions and remember this ordering. After this choice, the associated Hilbert space (or Fock
space) in which we can describe quantum states becomes a tensor product of spin $1/2$'s or qubits by the
identification
 \[
 |i_1,i_2,\cdots,i_N\rangle\equiv \left(\hat{a}_1^\dagger\right)^{i_1}
 \left(\hat{a}_2^\dagger\right)^{i_2}\cdots\left(\hat{a}_N^\dagger\right)^{i_N}
 |\Omega\rangle.
 \]
So it is indeed simple to represent a collection of $N$ fermions in a Hilbert space of $N$ spins \cite{Zanardi}.
When trying to simulate fermionic systems, one has to  work with the vector $C_{i_1,\cdots,i_N}$ which is defined
on this Hilbert space of spins.

The obvious question is now of course how fermionic Hamiltonians and creation/annihilation operators look like in
this spin representation. Let us e.g. consider the representation of the annihilation operator $\hat{a}_k$:
 \begin{eqnarray*}
 &&\langle\Omega|\left[\hat{a}_N^{j_N}\cdots\hat{a}_1^{j_1}\right]
 \hat{a}_k\left[\hat{a}_1^{\dagger i_1}\cdots\hat{ a}_N^{\dagger
 i_N}\right]|\Omega\rangle =\\
 &&\hspace{3cm} \delta_{j_k,0}\delta_{i_k,1} \prod_{n\ne k} \delta_{j_n,i_n}(-1)^{i_n}.
 \end{eqnarray*}

This can readily be derived by using the anticommutation relations. The effective Hamiltonian described in
spin-space can now easily be calculated. The transformation we did is exactly equivalent to the JWT which maps
commuting spin operators to anticommuting fermionic creation/annihilation operators. From this point of view, the
Jordan-Wigner transformation is not just a very useful trick but a natural consequence of the formalism of second
quantization. The Hamiltonian as defined on the Hilbert space of spins is thus obtained by carrying out a JWT on
the fermionic Hamiltonian.

Let us illustrate this transformation for some specific example. Due to superselection rules, physical
observables always contain an even number of creation/annihilation operators, and thus in the following we will
consider fermionic observables of this sort. Here we recall how typical operators transform (we neglected
constant terms):
 \begin{eqnarray*}
 \hat{a}_k^\dagger\hat{a}_k &\rightarrow &\sigma^z_k\\
 \hat{a}_k^\dagger \hat{a}_{k+n}+ c.c.&\rightarrow
 &\sigma^x_k S_{k,n} \sigma^x_{k+n}
 +\sigma^y_k S_{k,n} \sigma^y_{k+n}
 \end{eqnarray*}
where $S_{k,n}=\otimes_{l=k+1}^{k+n-1}\sigma^z_l$ is the string operator, which is the most distinctive feature
of this transformation and arises when considering hopping terms between distant fermions.

The central question of this paper is now the following: let us assume we have a system of fermions on some
lattice with a specific geometry; is it always possible to map a local Hamiltonian of fermions to a local
Hamiltonian of spins in this way? The answer to this question could clearly depend on the geometry and the type
of interactions. Let us first consider the 1-dimensional case; one immediately sees that one can always choose
the ordering of the fermions such that locality is preserved by the JWT. The situation seems to be very different
in 2 dimensions. Let us, e.g., consider a translational invariant model with nearest neighbor interactions and
hopping terms on a square lattice. Due to the string of $\sigma^z$ operators, there exists no ordering of the
fermions such that all local hopping terms in both horizontal and vertical direction map to local operators in
the spin representation. The same problem arises in all geometries different from a 1-dimensional chain. In the
following we will show how locality can nevertheless always be achieved by introducing extra degrees of freedom
in the form of extra fermions.

\begin{figure}[t]
  \centering
  \includegraphics[width=.4\linewidth]{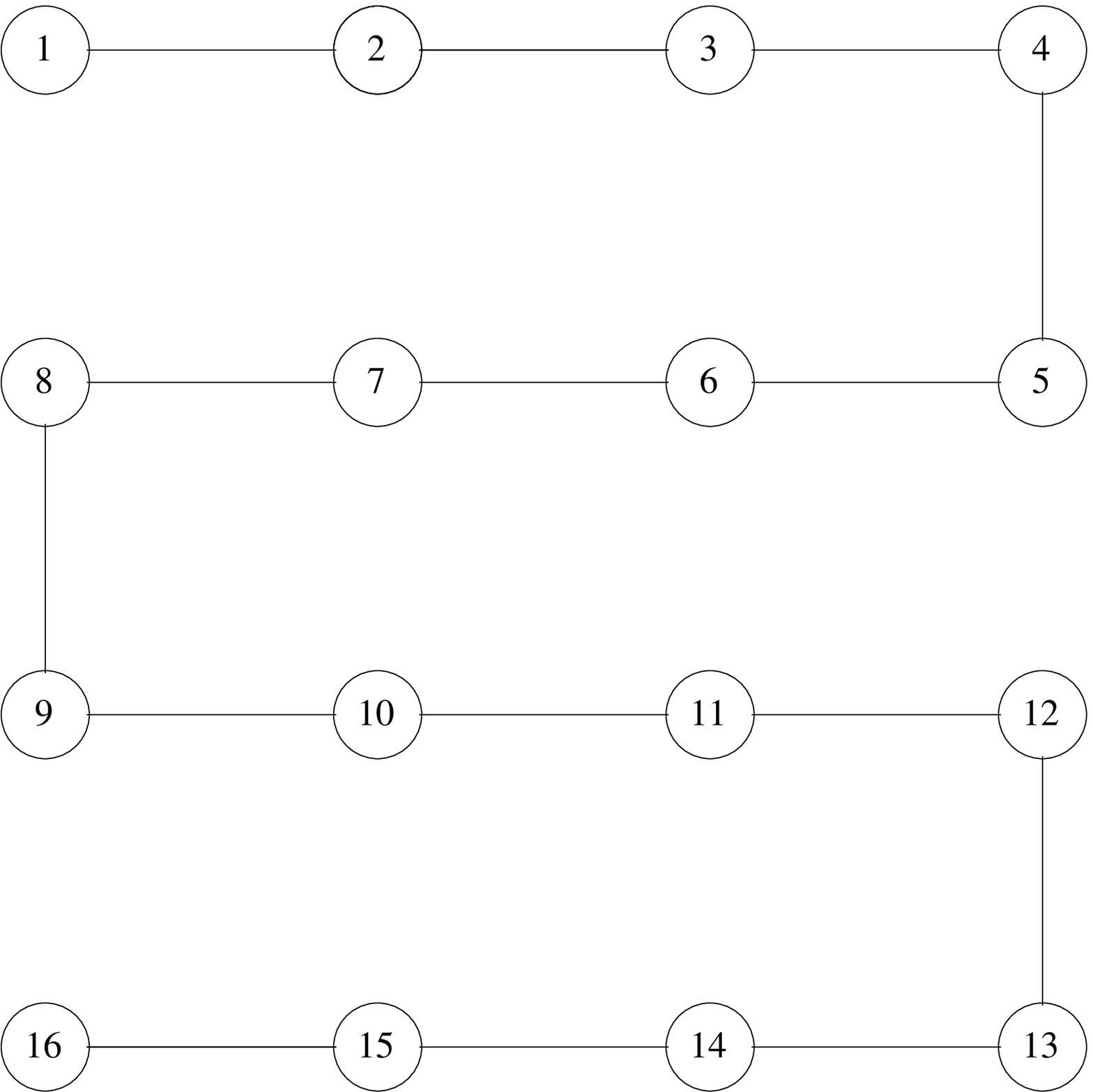}
  \caption{Enumerating fermions on a square 2-D lattice.}
  \label{fig1}
\end{figure}

\begin{figure}[t]
  \includegraphics[width=.7\linewidth]{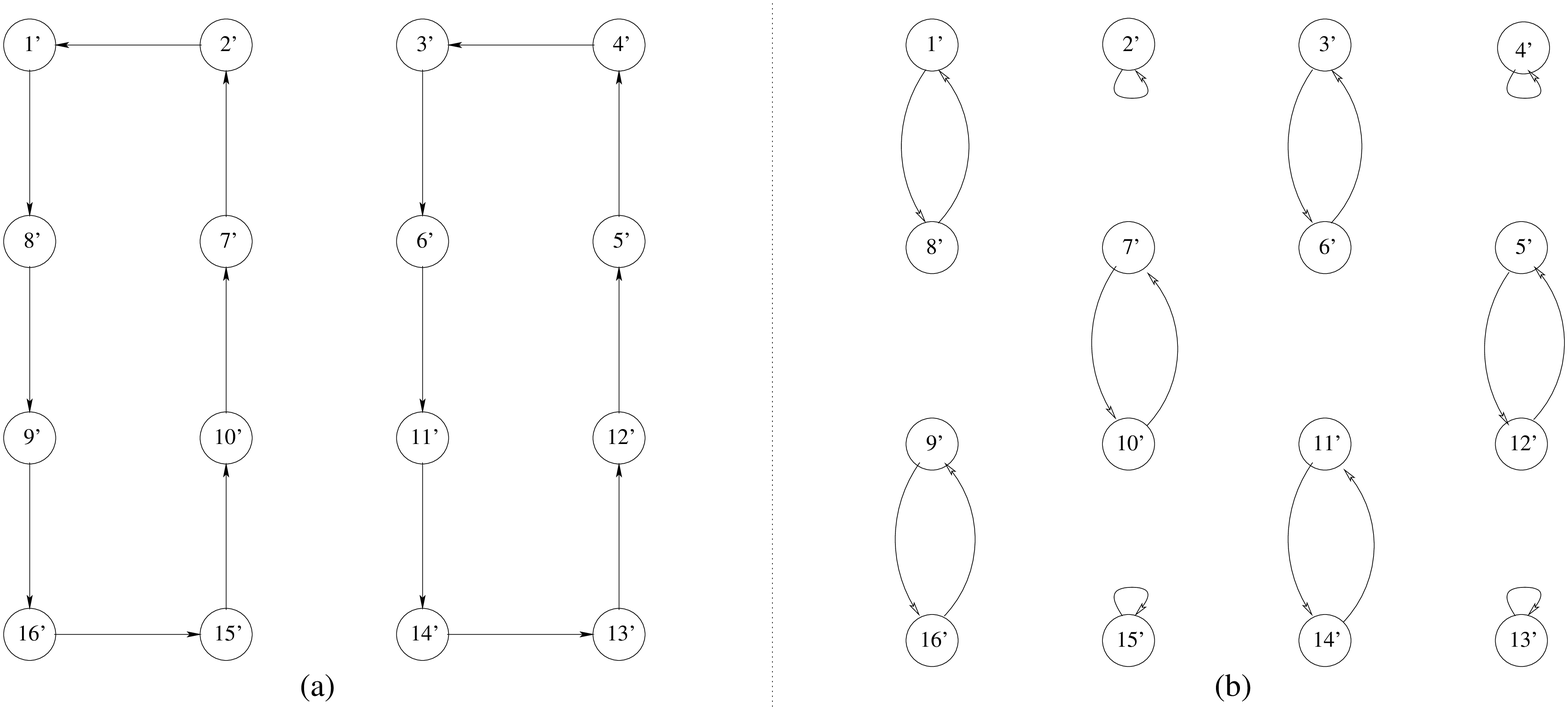}
  \caption{Two possible graphs specifying different Hamiltonians $\mathcal{H}_{aux}$ for the auxiliary fermions.}
  \label{error}
\end{figure}

For simplicity, let us consider a 2-dimensional square lattice of spinless fermions with $N$ columns and open
boundary conditions. We enumerate the fermions as in Fig. 1 and consider interaction terms of the form
\begin{eqnarray}
&&\hat{a}_i^\dagger \hat{a}_j\label{hopping}\\
&&\hat{a}_i^\dagger \hat{a}_i\hat{a}_j^\dagger \hat{a}_j\label{mang}
\end{eqnarray}
where either $i=j$, either $i$ and $j$ label two geometrically neighboring fermions. The JWT will map all
operators of the form (\ref{mang}) to local spin operators, but the hopping terms of the form (\ref{hopping})
will only result in local spin terms when $j=i\pm 1$ (i.e. all horizontal hopping terms and halve of the vertical
ones at the edges). So there is clearly only a problem with the vertical hopping terms.

The idea is now to introduce an extra fermion $\hat{b}_i$ for each mode $\hat{a}_i$, and impose that these extra
fermions are in the ground state $|\chi\rangle$ of the Hamiltonian
 \[
 \mathcal{H}_{aux}=\sum_{\{i,j\}}\hat{P}_{ij}\equiv \sum_{\{i,j\}}
 -(\hat{b}_i+\hat{b}_i^\dagger)(\hat{b}_j-\hat{b}_j^\dagger),
 \]
where the set $\{i,j\}$ includes only pairs $(i,j)$ that correspond to the directed edges in Fig. 2(a) from
vertex $i$ to $j$ \footnote{Note that if the number of columns $N$ is odd, it is necessary to take an auxiliary
lattice with $N-1$ columns such as to have an even number of columns to close the loops in Fig. 2a.}. It can
readily be verified that all the $\hat{P}_{ij}$ commute, $\hat{P}_{ij}^2=\openone$, and that the gap of the
Hamiltonian $\mathcal{H}_{aux}$ is $2$. Its unique ground state has the property that
\[\forall (i,j)\in\{i,j\}, \hspace{1cm}\hat{P}_{ij}|\chi\rangle=|\chi\rangle.\]
Actually, we can define the operators $\hat{c}_i=(\hat{b}_i+\hat{b}_i^\dagger)$ and
$\hat{d}_i=-\imath(\hat{b}_i-\hat{b}_i^\dagger)$ as two Majorana fermions making up the usual fermions
\cite{Bravyi}.

Next we consider the system consisting of the original fermions $\hat{a}_i$ and of the auxiliary system together;
geometrically, we assume that each site is occupied by two fermions, the original $\hat{a}_i$ and the new one
$\hat{b}_{i'}$. We leave the Hamiltonian on the auxiliary modes $\mathcal{H}_{aux}$ unchanged, but alter the
vertical hopping terms of the physical Hamiltonian in the following way:
 \[ \hat{a}_i^\dagger \hat{a}_j\rightarrow \hat{a}_i^\dagger \hat{a}_j
 \hat{P}_{i'j'}=\hat{a}_i^\dagger \hat{a}_j \left(\imath
 \hat{c}_{i'}\hat{d}_{j'}\right).
 \]
Clearly, all these vertical hopping terms commute with $\mathcal{H}_{aux}$, and hence the $\hat{P}_{ij}$ are
constants of motion (i.e. $+1$), and as such the ground state of the full system will be the tensor product of
the original ground state and $|\chi\rangle$ \footnote{Strictly speaking, this is only true if the energy gap in
the auxiliary system is greater than the energy that could potentially be gained by attributing negative weights
to vertical hopping terms; this can always be guaranteed by multiplying the auxiliary Hamiltonian by a big enough
factor.}. But the advantage of the new formulation is immediately clear when the fermions are reordered in the
way $1,1',2,2',...$ when doing second quantization. As a result, the hopping terms in the vertical direction will
now correspond to quartic interaction terms which will remain local by doing a JW transformation: the string
operator of $k'$ will cancel the one of $k$. Note that the configuration of the auxiliary fermions shown in Fig.
2(a) is only one possible choice and many other loop structures would lead to similar results (one possibility
would e.g. be to have only loops with 2 vertical edges such as in Fig 2(b)).

But the problem is not quite solved yet, because the JW-transformation will turn $\mathcal{H}_{aux}$ into a
nonlocal Hamiltonian. Note however that the JW does not map the vertical hopping at the alternating edges of the
lattice hopping into nonlocal ones. Consider e.g. the pair $(N,N+1)$ and associated $\hat{P}_{N,N+1}$  at such an
edge (here $N$ is the number of columns; $N=4$ in Fig. 1); let us now make the following substitution in
$\mathcal{H}_{aux}$:
 \begin{eqnarray*}
 \hat{P}_{N-1,N+2}&\rightarrow&\hat{P}_{N-1,N+2}\hat{P}_{N,N+1}\\
 \hat{P}_{N-2,N+3}&\rightarrow&\hat{P}_{N-2,N+3}\hat{P}_{N-1,N+2}\\
 &\vdots&\\
 \hat{P}_{1,2N}&\rightarrow&\hat{P}_{1,2N}\hat{P}_{2,2N-1}
 \end{eqnarray*}
and similar for all other rows. Due to the fact that all $\hat{P}$ commute, this substitution does not affect the
ground state and the gap of $\mathcal{H}_{aux}$. The advantage of using this new Hamiltonian is however obvious:
the JW-transformation will map it to a local Hamiltonian with local plaquette terms: the string of
$\sigma^z$-operators generated by $\hat{c}_{i'}$ will be cancelled by the ones of $\hat{c}_{(i+1)'}$. So we
indeed managed to transform a local Hamiltonian of fermions into a local Hamiltonian of spins! One could argue
that this indicates that fermionic systems potentially represent the low-energy sector of bosonic systems (see
e.g. \cite{Wen1}).

Let us illustrate the kind of terms that will appear in the spin Hamiltonian for the case of the square $N\times
N$ lattice of spinless fermions. Without loss of generality, we assume $N$ to be even. We arrange the auxiliary
fermions as in Fig. 2(a), and imagine that we have two superimposed layers  of $N\times N$ qubits, the first
layer corresponding to the \emph{physical} qubits and the second one to the auxilliary ones. We label the Pauli
spin operators acting on the first layer as $X_{k,l},Y_{k,l},Z_{k,l}$ and on the second layer
$\tilde{X}_{k,l},\tilde{Y}_{k,l},\tilde{Z}_{k,l}$ with $1\leq k\leq N$ labelling the column and $l$ the row. Let
us systematically show how all terms in a fermionic will look like in the spin picture.

First of all we observe that the following terms transform trivially:

\begin{eqnarray*}
\hat{a}_{k,l}^\dagger\hat{a}_{k,l}&\rightarrow&Z_{k,l}-\openone\\
\left(\hat{a}_{k,l}^\dagger\hat{a}_{k,l}\right)\left(\hat{a}_{k',l'}^\dagger\hat{a}_{k',l'}\right)&\rightarrow&
\left(Z_{k,l}-\openone\right)\left(Z_{k',l'}-\openone\right)
\end{eqnarray*}

Horizontal hopping terms transform as
\[\hat{a}_{k,l}^\dagger \hat{a}_{k+1,l}+\hat{a}_{k+1,l}^\dagger \hat{a}_{k,l}\rightarrow
\left(X_{k,l}X_{k+1,l}+Y_{k,l}Y_{k+1,l}\right)\tilde{Z}_{k,l}\]

Vertical hopping terms become
\begin{eqnarray*}
\hat{a}_{k,l}^\dagger \hat{a}_{k,l+1}+\hat{a}_{k,l+1}^\dagger \hat{a}_{k,l}&\rightarrow & \left(X_{k,l}X_{k,l+1}+Y_{k,l}Y_{k,l+1}\right)(-1)^{l+1}\left(\tilde{X}_{k,l}\tilde{Y}_{k,l+1}\right)\hspace{.3cm} \rm{(k=odd)}\\
&\rightarrow &
\left(X_{k,l}X_{k,l+1}+Y_{k,l}Y_{k,l+1}\right)(-1)^{l+1}\left(\tilde{Y}_{k,l}\tilde{X}_{k,l+1}\right)\hspace{.3cm}
\rm{(k=even)}
\end{eqnarray*}
depending on whether $k$ is even or odd.

The gauge conditions can be imposed by including the following 6-body local terms in the spin Hamiltonian:
\begin{eqnarray*}
\sum_{\rm{l=odd,k=odd}}
\left(Z_{k+1,l}Z_{k,l+1}\right)\left(\tilde{Y}_{k,l}\tilde{Y}_{k+1,l}\tilde{Y}_{k,l+1}\tilde{Y}_{k+1,l+1}\right)\\
\sum_{\rm{l=odd,k=even}}
\left(Z_{k+1,l}Z_{k,l+1}\right)\left(\tilde{X}_{k,l}\tilde{X}_{k+1,l}\tilde{X}_{k,l+1}\tilde{X}_{k+1,l+1}\right)\\
\sum_{\rm{l=even,k=odd}}
\left(Z_{k,l}Z_{k+1,l+1}\right)\left(\tilde{Y}_{k,l}\tilde{Y}_{k+1,l}\tilde{Y}_{k,l+1}\tilde{Y}_{k+1,l+1}\right)\\
\sum_{\rm{l=even,k=even}}
\left(Z_{k,l}Z_{k+1,l+1}\right)\left(\tilde{X}_{k,l}\tilde{X}_{k+1,l}\tilde{X}_{k,l+1}\tilde{X}_{k+1,l+1}\right)
\end{eqnarray*}
It can easily be checked that these terms commute and also commute with the hopping terms. Finally, boundary
conditions are required to fix the gauge uniquely (note that an ordering of the fermions and the corresponding
JWT was crucial to make these terms local):
\begin{eqnarray*}
\sum_{\rm{k=odd}}\tilde{X}_{k,1}Z_{k+1,1}\tilde{X}_{k+1,1}&&\\
\sum_{\rm{k=odd}}\tilde{X}_{k,N}Z_{k,N}\tilde{X}_{k+1,N}&&\\
\sum_{\rm{k=odd}}\tilde{X}_{N,k}Z_{N,k+1}\tilde{X}_{N,k+1}&&\\
\sum_{\rm{k=even}}\tilde{Y}_{1,k}Z_{1,k+1}\tilde{Y}_{1,k+1}&&\\
\end{eqnarray*}

The sum of all those terms (of course with the appropriate prefactors of the original model) is the local spin
Hamiltonian whose low energy sector is equivalent to the original fermionic model. Let's try to get some insight
into all these terms. The hopping terms give rise to the well known $XX+YY$ interactions, but these are mediated
by a $\tilde{Z}$ term in the horizontal case and a $\tilde{X}\tilde{Y}$ term in the vertical case; it is
intriguing how the noncommuting nature of these extra terms should be related to the topological features of
fermionic systems. Let us now have a closer look at the gauge conditions. Let us first assume that all spins in
the first layer are completely polarized in the $Z$-direction; this would correspond to a completely trivial
Hamiltonian for the original fermions. The qubits in the second layer would then interact via plaquette
interactions of the form $XXXX$ or $YYYY$, as all $ZZ$ expectation values are equal to $1$. It can readily be
checked that the effective Hamiltonian acting on the second layer is then exactly equal to the celebrated toric
code Hamiltonian of Kitaev \cite{Toric} if one rotates the lattice over $\pi/4$: the topological features of the
toric code states seem to be exactly the ones needed to mediate the desired mapping from local interactions of
fermions to spins. It would be very interesting to investigate this connection between the topological features
of toric code states and emergent fermions further, and to arrive at a more intuitive explanation for the
proposed mathematical construction.

This construction can readily be generalized to fermions with spin and any regular lattice geometry; in e.g. a
3-D cubic lattice, one would need two extra fermions per site (i.e. 4 Majorana fermions). The Hamiltonian
characterizing the auxiliary fermions can in this 3-D case still be made local by multiplying four commuting
$c_id_j$-terms with each other. A triangular lattice can e.g. be obtained by embedding it into a square lattice;
in this particular case some fermions would not be coupled to any other ones, and extra Majorana fermions guiding
extra horizontal hopping terms have to be introduced. One can also include long-range hopping terms by
multiplying them with multiple factors $\hat{P}_{ij}$. A 2-D square lattice of fermions with an additional spin
$1/2$ degree of freedom could either be treated by considering two layers of the spinless system, or by working
with spin $3/2$ objects instead of qubits.

An interesting problem arises when the same analysis is carried out on systems with periodic boundary conditions.
Let us e.g. consider the case of a 1-D ring; here we can add two Majorana fermions to each original fermion, and
use the link between the first unpaired Majorana fermion and the last one as a tool to make the original
interaction term between the first and last fermion quartic. However, the JW-transformation will make the
interaction term between these two Majorana fermions nonlocal, and there does not seem to be a way around it: the
topology of the ring enforces us to have one nonlocal term in the final spin Hamiltonian \footnote{Note that this
nonlocal term does not necessarily present a problem for the numerical simulation of these systems. In e.g. the
XY-chain, one just has to solve the problem in the sector with even and odd number of particles independenty.}.
Similarly, in the case of a cylinder one nonlocal term will appear, and in the case of a torus two nonlocal
terms. In other words, we can speculate that the ground state of the corresponding spin system has the property
that it cannot be the unique ground state of a local spin Hamiltonian: the associated local Hamiltonian (i.e.
without nonlocal term) will possibly have a two-fold degenerate ground state, and a nonlocal operator which is
topologically nontrivial will be needed to specify the sector.  This phenomenon has indeed been identified in the
case of the toric code states \cite{Toric} of Kitaev, and is related to topological quantum order \cite{Wen2}.

\begin{figure}[t]
  \includegraphics[width=.6\linewidth]{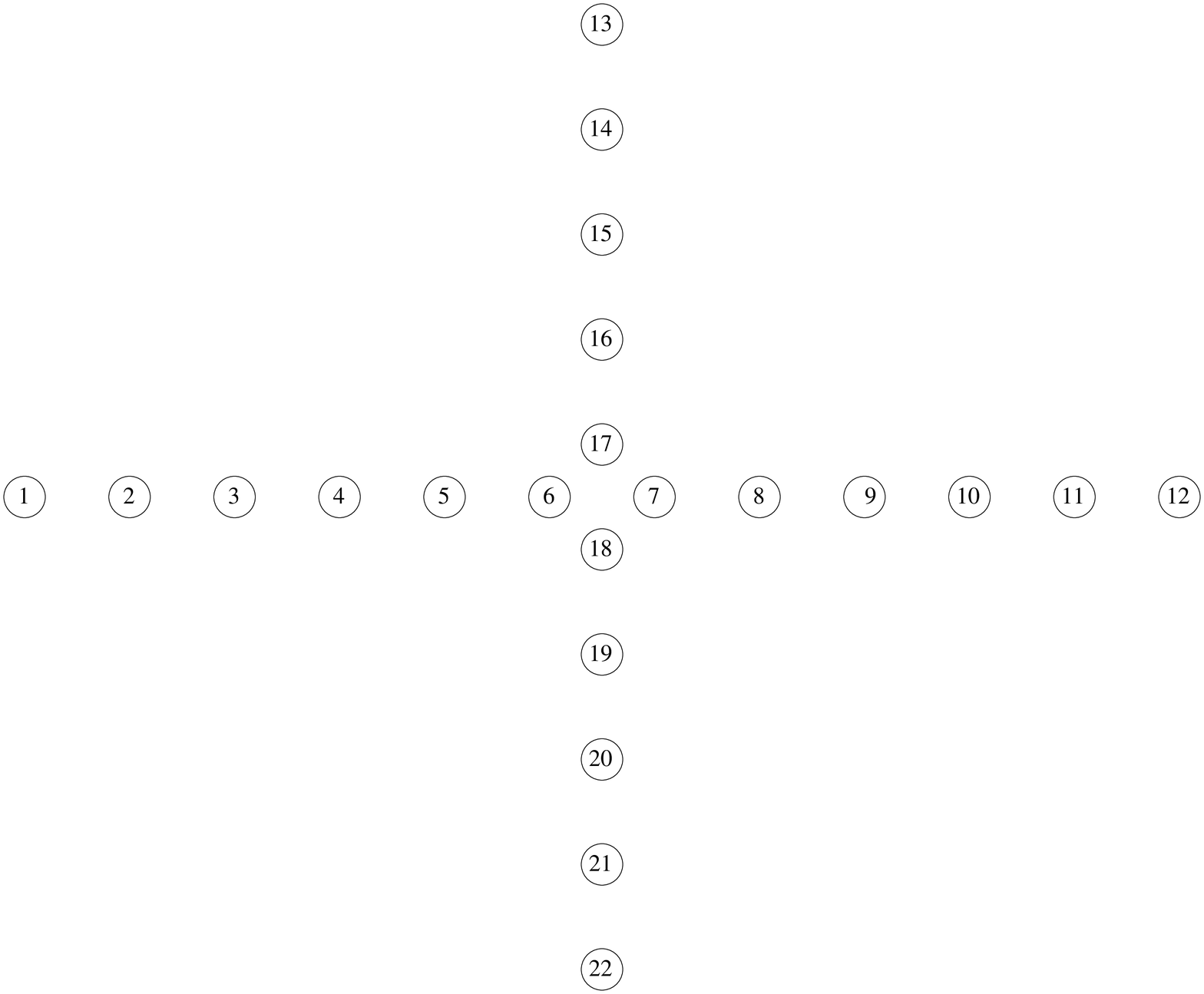}
  \caption{Enumerating fermions in the case of a 2-channel Kondo-like impurity problem; 1-5 represent the spin up fermions of the first channel, 6-7 the impurity, 8-12 the spin down fermions of the first
  channel, and 13-17 and 18-22 respectively the spin up and down fermions of the second channel.}
  \label{error}
\end{figure}

Instead of regular lattices, one can also ask the question whether a similar mapping is possible in the case of a
star-like geometry which e.g. occurs in multichannel Kondo-related impurity problems \cite{Wilson,VW05}. Let us
therefore consider an impurity coupled to two different kind of fermions each having an extra spin degree of
freedom. Enumerating  the different fermions as in Fig. 3 (here fermions nr. 6 and 7 correspond to the impurity),
one finds that again a long range string operator will appear between the second type of fermions and the
impurity (i.e. between $\{6,17\}$ and $\{7,18\}$). As in the previous case, we introduce two pairs of Majorana
fermions and multiply the problematic hopping terms with the appropriate Majorana operators. This will indeed
make the hopping terms local after the JWT. The problem in this case however is that after the JW there is no way
to impose the identities $ic_kd_k=+1$ for both pairs in a local way. But here one can easily see that there is no
need to impose that constraint: the effective Hamiltonian after the JW is block diagonal, one sector
corresponding to $ic_kd_k=+1$ and the other one to $ic_kd_k=-1$; the second block is equivalent to the first one
under local unitary operations. Indeed, the local transformation $a_i\rightarrow \exp(i\pi a_i^\dagger a_i)a_i$
on all fermions of one type will leave the interaction terms between them invariant, and will only lead to a
negative hopping term when coupling them to the impurity. So the ground states in the two sectors are related by
local unitaries: no gauge fixing interaction term needs to be imposed, and when doing simulations on the spin
Hamiltonian one can always check the sector one ends up in. So it is again enough to diagonalize a local spin
Hamiltonian.

In conclusion, we have shown how it is possible to map a local problem of fermions onto a local problem of spins
by using auxiliary Majorana fermions. This leads to a 2-dimensional version of the Jordan Wigner transformation.
We believe that, apart from its fundamental interest, this map may become very useful in the context of numerical
description of fermionic systems, since in that case local spin Hamiltonians seem to be better approximated. An
open question is whether one can use this mapping to solve certain interesting spin Hamiltonians exactly by
transforming it back into free fermions, as it is done in 1-dimensional investigations and in the recent work of
Kitaev \cite{Kitaev}.

We thank S. Bravyi, J. Preskill and G. Vidal for discussions. After completion of this work, we became aware of
the work by Ball \cite{Ball}, and we thank X. Wen for bringing this work to our attention. Work supported by the
Gordon and Betty Moore Foundation, European projects CONQUEST, TOPQIP and QUPRODIS, and the DFG (SFB 631).

\end{document}